\documentclass[12pt]{article}

\usepackage{amssymb, amsmath}
\usepackage{graphicx}
\usepackage{cite}
\def\be{\begin{equation}}
\def\ee{\end{equation}}
\def\bea{\begin{eqnarray}}
\def\eea{\end{eqnarray}}
\def\tr{ \text{Tr}}
\def\td{ \textrm{d}}

\title{ {\bf Uniqueness of extreme horizons in Einstein-Yang-Mills theory}}

\author{Carmen Li\footnote{K.K.Li@sms.ed.ac.uk} \  and James Lucietti\footnote{j.lucietti@ed.ac.uk } \\ \\  \small \sl  School of Mathematics and Maxwell Institute for Mathematical Sciences, \\ \small \sl  University of Edinburgh, King's Buildings, Edinburgh, EH9 3JZ, UK}

\date{}

\begin{document}

\maketitle

\begin{abstract}
We consider stationary extreme black hole solutions to the Einstein-Yang-Mills equations in four dimensions, allowing for a negative cosmological constant. We prove that any axisymmetric black hole of this kind possesses a near-horizon AdS$_2$ symmetry and deduce its near-horizon geometry must be that of the abelian embedded extreme Kerr-Newman (AdS) black hole. We also show that the near-horizon geometry of any static black hole is a direct product of AdS$_2$ and a constant curvature space. 
\end{abstract}

\section{Introduction}
Extreme black holes are important in studies of quantum gravity since they possess zero temperature. A key geometric structure which all known examples possess is a near-horizon AdS$_2$ symmetry. This symmetry has played a fundamental role in developing various quantum descriptions of extreme black holes, see e.g.~\cite{Strominger:1998yg, Sen:2008vm}. It has even lead to the proposal that the extreme Kerr black hole is described by a two-dimensional CFT~\cite{Guica:2008mu}. 

The AdS$_2$ near-horizon symmetry has been established in a wider context via near-horizon symmetry enhancement theorems for $D=4,5$ extreme black holes~\cite{Kunduri:2007vf} and also for $D>5$~\cite{Figueras:2008qh, Lucietti:2012sa}, under various assumptions regarding the rotational symmetry. In $D=4,5$ the theorem is valid in a general class of theories of Einstein gravity  coupled to an arbitrary number of Maxwell fields and uncharged scalars (with a non-positive potential)~\cite{Kunduri:2007vf}. This includes a number of consistent truncations of higher dimensional supergravity theories, such as $D=4,5$ minimal (gauged) supergravity coupled to vector multiplets. Typically, these are special cases of more general consistent truncations such as $D=4,5$ maximal (gauged) supergravity, which contain more general types of matter such as charged scalar fields and non-abelian gauge fields. It is therefore of interest to investigate whether the near-horizon symmetry enhancement phenomenon persists in the presence of such fields. In this note we will focus on four dimensional extreme black holes with {\it non-abelian gauge fields}. 

It has been known for sometime that the four dimensional black hole uniqueness theorems fail in the presence of non-abelian gauge fields, see~\cite{Volkov:1998cc} for a review. Most strikingly Einstein-Yang-Mills theory admits an infinite number of asymptotically flat, static and spherically symmetric solutions, including both smooth solitons~\cite{Bartnik:1988am} and regular black holes~\cite{Volkov:1989fi, Kuenzle:1990is, Bizon:1990sr}. These were first found numerically and subsequently their existence was established rigourously~\cite{Smoller:1993pe, Smoller:1993bs}. In fact many of the components of the black hole uniqueness theorems do not work when coupled to a non-abelian gauge field~\cite{Heusler:1996ft, Chrusciel:2012jk}. Non-rotating black holes need not be static, static ones need not be spherically symmetric, and as already mentioned even spherically symmetric ones are not unique. On the other hand, the rigidity theorem still applies, which guarantees that a rotating black hole must be axisymmetric. However, the Einstein equations for stationary and axisymmetric spacetimes do not guarantee orthogonal transitivity of the isometry group (as in Einstein-Maxwell theory). Hence the Weyl-Papapetrou form for the metric is overly restrictive, and furthermore, even assuming this does not lead to an integrable 2d theory (as in Einstein-Maxwell).

Interestingly, if the gauge group is $SU(2)$, four-dimensional Einstein-Yang-Mills theory with a negative cosmological constant is a consistent truncation of 11d supergravity on a squashed $S^7$~\cite{Pope:1985bu}. It is worth noting that  in this context non-abelian Anti de Sitter black hole solutions also exist~\cite{Winstanley:1998sn,Winstanley:2008ac }.\footnote{Of course, in the presence of a cosmological constant even the (electro-)vacuum black hole uniqueness theorems are not valid.} Such solutions are of interest in the context of the AdS/CFT dualities and in the case of planar horizons have been used to model phase transitions analogous to superfluidity/superconductivity~\cite{Gubser:2008zu}.

Most of the investigations of non-abelian black holes have focused on non-extreme black holes. Given the importance of non-abelian equilibrium states in quantum theory it is natural to ask: does the gross violation of uniqueness persist for {\it extreme} black holes? It appears this question has not been fully investigated even in four dimensions.  A result in this direction suggesting this is not the case is that static and spherically symmetric extreme black holes to $SU(2)$-Einstein-Yang-Mills theory are uniquely given by the abelian embedding of the Reissner-Norstr\"om black hole~\cite{Smoller:1996hb, Galtsov:1989ip, Bizon:1992pi}. A natural method for investigating the question more generally is to attempt to classify {\it near-horizon geometries} of non-abelian extreme black holes.   In fact for {\it static} black holes this has been already considered under certain restrictive assumptions~\cite{Hajicek:1982ki, Bicak:1994tq}. Thus, our main focus will be stationary (non-static) black holes. The analogous problem in Einstein-Maxwell theory,  including a cosmological constant,  has been previously solved~\cite{Lewandowski:2002ua, Kunduri:2008rs, Kunduri:2008tk}.

The first hurdle is that the AdS$_2$ near-horizon symmetry theorems mentioned above, do not immediately apply in the presence of non-abelian gauge fields. In fact recently it was shown that the enhancement of symmetry of the near-horizon geometry follows from orthogonal transitivity of stationary and axisymmetric solutions~\cite{Lucietti:2012sa}. As mentioned above, the Einstein-Yang-Mills equations do not imply orthogonal transitivity (unlike in Einstein-Maxwell theory), thus raising the question: are there non-abelian near-horizon geometries without an AdS$_2$ symmetry?

In this note we will solve this problem within the simplest set up: four dimensional Einstein-Yang-Mills with a compact semi-simple gauge group and a cosmological constant $\Lambda$ (mainly focusing on $\Lambda \leq 0$). We show that in fact the AdS$_2$ symmetry theorem can be generalised to axisymmetric near-horizon geometries with cross-sections of the horizon of spherical topology. This requires an extra global argument as compared to the Einstein-Maxwell case~\cite{Kunduri:2008tk}. Given this, the system of equations is then essentially equivalent to the Einstein-Maxwell case, allowing us to show that the most general solution of this kind is the near-horizon geometry of the abelian embedded extreme Kerr-Newman black hole (with cosmological constant). 

We also show that there are no {\it non-static} axisymmetric near-horizon geometries with toroidal  cross-sections of the horizon. This is also the case in Einstein-Maxwell theory, a fact that does not seem to have been shown before for $\Lambda<0$.\footnote{For $\Lambda \geq 0$ this fact immediately follows by integrating the horizon scalar curvature~\cite{Kunduri:2008tk} .}  For completeness, by following the method used for Einstein-Maxwell theory we completely classify static near-horizon geometries, revealing that the only solutions with a compact horizon section are direct products of AdS$_2$ and a constant curvature space.

\section{Non-abelian gauge fields near an extreme horizon}

Let $(M, g_{\mu\nu})$ be a four-dimensional spacetime satisfying the Einstein-Yang-Mills equations with a cosmological constant $\Lambda$. We will assume the gauge group is a compact Lie group whose Lie algebra $\mathfrak{g}$ is semisimple. Hence $\mathfrak{g}$ admits a positive definite invariant metric $( \cdot, \cdot )$ which we will denote by $\tr(A B) \equiv ( A, B)$ where $A,B \in \mathfrak{g}$ (i.e. the Killing form).  

We denote the $\mathfrak{g}$-valued Yang-Mills gauge field by $\mathcal{A}_\mu$ and the the gauge-covariant derivative of any $\mathfrak{g}$-valued differential form is $\mathcal{D}X = \td X+[\mathcal{A}, X ]$. The Yang-Mills field strength is $\mathcal{F}= \td \mathcal{A} + \tfrac{1}{2} [\mathcal{A}, \mathcal{A} ]$ and the Bianchi identity is $\mathcal{D} \mathcal{F}=0$.  Gauge transformations act as $X \mapsto U X U^{-1}$ and $\mathcal{A} \mapsto U \mathcal{A} U^{-1}- \td U U^{-1}$ where $U$ is a group-valued function. 

The Einstein-Yang-Mills equations are then
\bea
R_{\mu \nu} &=& 2\, \tr \left( {\mathcal{F}_\mu}^\delta \mathcal{F}_{\nu\delta} - \tfrac{1}{4} g_{\mu \nu} \mathcal{F}_{\rho \sigma} \mathcal{F}^{\rho \sigma} \right) + \Lambda g_{\mu \nu} \label{einsteinequation} \\
\mathcal{D} \star \mathcal{F} &=& 0 \label{ymfieldeqn}
\eea
where $\star$ denotes the Hodge dual with respect to $g_{\mu\nu}$ .

We will consider smooth solutions $(g_{\mu \nu}, \mathcal{A}_\mu)$ which are invariant -- up to gauge transformations -- under some symmetry group. Let us recall various well known general facts for such solutions~\cite{Forgacs:1979zs, Heusler:1996ft}. Explicitly,  if $\xi^\mu$ is a Killing vector field of $g_{\mu\nu}$ then $\mathcal{L}_\xi \mathcal{A} = \mathcal{D} \mathcal{V}_\xi$ where $\mathcal{V}_\xi$ is a $\mathfrak{g}$-valued function. This condition is gauge covariant provided gauge transformations act as $\mathcal{V}_\xi \mapsto U \mathcal{V}_\xi U^{-1}- (\mathcal{L}_\xi U)U^{-1}$. It follows that $\mathcal{L}_\xi \mathcal{F} = [ \mathcal{F}, \mathcal{V}_\xi ]$; hence for non-abelian fields there is no gauge-invariant notion of an invariant field strength. It is convenient to introduce the ``electric" 1-form $\mathcal{E}= -i_\xi \mathcal{F}$; it is then easy to show that there exists a $\mathfrak{g}$-valued potential $\mathcal{W} = i_\xi \mathcal{A} - \mathcal{V}_\xi$ such that $\mathcal{E}=  \mathcal{D} \mathcal{W}$. Observe that $\mathcal{D} \mathcal{E} = [ \mathcal{F}, \mathcal{W}]$. One can also introduce a ``magnetic" 1-form $\mathcal{B}= i_\xi \star \mathcal{F}$; for a non-abelian field there is no associated potential, although by contracting the Yang-Mills equation with $\xi$ one can show $\mathcal{D} \mathcal{B} = [\mathcal{W}, \star \mathcal{F}]$.

We are now ready to introduce our setup. Suppose $(M,g_{\mu\nu})$ contains a smooth {\it degenerate} Killing horizon $\mathcal{N}$ of a complete Killing vector field $K^\mu$, with a compact cross-section $H$ (i.e. a 2-dimensional submanifold of $\mathcal{N}$ intersected by each orbit of $K$ exactly once). Let $U^{\mu}$ be tangent to the null geodesics which are orthogonal to $H$ and satisfy $K \cdot U=1$. In the neighbourhood of such a horizon one can define Gaussian null coordinates $(v,r,x^1,x^2)$, so that $K= \partial /\partial v$, $U=\partial /\partial r$, where $r=0$ is the horizon $\mathcal{N}$, and  $(x^1,x^2)$ are arbitrary coordinates on $H$ (which corresponds to $r=0$ and $v=\text{const}$). The metric in these coordinates takes the form~\cite{isenberg}
\be
g_{\mu\nu} \td x^\mu \td x^\nu = 2\, \td v \left( \td r + r h_a(r,x) \td x^a + \tfrac{1}{2} r^2 F(r,x)\td v \right) + \gamma_{ab}(r,x) \td x^a \td x^b
\ee
where $F,h_a, \gamma_{ab}$ are all smooth functions.  Degeneracy of the horizon corresponds to $g_{vv}={\cal O}(r^2)$.

We will assume that the Killing field $K$ leaves the gauge field invariant up to gauge transformations, i.e. $\mathcal{L}_K \mathcal{A} = \mathcal{D} \mathcal{V}_K$, and denote the associated potential  defined above by $\mathcal{W}= i_K \mathcal{A} - \mathcal{V}_K$. Now, for any Killing horizon one must have $R_{\mu\nu} K^\mu K^\nu|_{\mathcal{N}}=0$. On the other hand the Einstein equations (\ref{einsteinequation}) imply  $R_{\mu\nu} K^\mu K^\nu|_{\mathcal{N}}= \tr ( \mathcal{E}_\mu \mathcal{E}^\mu + \mathcal{B}_\mu \mathcal{B}^\mu )|_{\mathcal{N}}$. It follows that in Gaussian null coordinates $\mathcal{E}_{a}|_{r=0}=\mathcal{B}_a|_{r=0}=0$. We will now recast these as equations on $H$. 

Let us denote the restriction of any quantity to $H$ by a ``hat". In particular we write the gauge field on $H$ as $\hat{\mathcal{A}}= \hat{\mathcal{A}}_a \td x^a$  and the corresponding Yang-Mills field strength on $H$ is $\hat{\mathcal{F}} \equiv \hat{\td} \hat{\mathcal{A}} + \tfrac{1}{2} [ \hat{\mathcal{A}}, \hat{\mathcal{A}} ]$. Also let $\hat{\mathcal{D}} \equiv \hat{\td} \cdot  + [ \hat{\mathcal{A}}, \cdot]$ be the gauge-covariant derivative on $H$. The condition $\mathcal{E}_{a}|_{r=0}=0$ then implies the equation on $H$
\be
\label{DW}
\hat{\mathcal{D}}_a \hat{\mathcal{W}}=0  \; .
\ee 
 We deduce that $\hat{\mathcal{D}}^2 \hat{\mathcal{W}} = [ \hat{\mathcal{F}}, \hat{\mathcal{W}}]=0$. On the other hand, by contracting  $\mathcal{D} \mathcal{B} = [\mathcal{W}, \star \mathcal{F}]$ with vector fields tangent to $H$, the condition $\mathcal{B}_a|_{r=0}=0$ implies that $[ \hat{\mathcal{W}}, \hat{\mathcal{F}}_{vr} ] =0$ on $H$. 

By a gauge transformation  we may set $\mathcal{V}_K=0$, so that $\mathcal{L}_K \mathcal{A}=0$; it follows that $\mathcal{L}_K \mathcal{F}=0$ and the electric potential $\mathcal{W} = i_K \mathcal{A}$. In Gaussian null coordinates the components of the gauge field and field strength are now both $v$-independent: $\partial_v \mathcal{A}_\mu=\partial_v \mathcal{F}_{\mu\nu}=0$. There is a residual gauge freedom which includes any gauge transformation satisfying $\partial_v U=0$: using this we can  further fix the gauge $\mathcal{A}_r = 0$.
 In this gauge, the most general gauge field is thus
\be
\mathcal{A} = \mathcal{W}(r,x) \td v +{\mathcal{A}}_a(r,x) \td x^a  \; .
\ee
Residual gauge transformations now satisfy $\partial_v U= \partial_r U=0$.  

The remaining gauge field data on $H$ is therefore explicitly given by $\hat{\mathcal{A}}_a = \mathcal{A}_a|_{r=0}$ and $\hat{\mathcal{W}} = \mathcal{W}|_{r=0}$. For convenience we also define $\hat{E} \equiv \partial_r \mathcal{W} |_{r=0}$ and $\hat{G} \equiv \hat{\star}_2 \hat{\mathcal{F}}$, which are $\mathfrak{g}$-valued functions on $H$. From above it follows that
\be
[ \hat{\mathcal{W}}, \hat{E} ] =0 \qquad  [ \hat{\mathcal{W}}, \hat{G} ]=0  \; ,  \label{Wcomm}
\ee
where we have used $\mathcal{F}_{vr}= -\partial_r \mathcal{W}$. 

We wish to investigate the constraints imposed by the Einstein-Yang-Mills equations on the horizon geometry. Usually, a convenient way to do this is to consider the field equations for the {\it near-horizon geometry}. This is defined by taking the near-horizon limit, which consists of first performing the diffeomorphism $(v,r) \to ( v/ \epsilon, \epsilon r)$ and then taking the limit $\epsilon \to 0$~\cite{Kunduri:2007vf}. Due to the degeneracy of the horizon this limit always exists for the metric. As we will show below, in our gauge, the limit also always exists for the Yang-Mills field strength $\mathcal{F}$. However, if $\hat{\mathcal{W}} \neq 0$ this limit does not exist for the gauge field $\mathcal{A}$. Since $\mathcal{A}$ appears explicitly in the Yang-Mills equation (i.e not just through $\mathcal{F}$), it is therefore not clear if one can take the near-horizon limit of this equation (for the Einstein equation this is not an issue). Therefore we will not immediately take the near-horizon limit, but instead expand the Einstein-Yang-Mills equations for the full spacetime fields for small values of the affine parameter $r$, i.e. near the horizon $\mathcal{N}$.

First we consider the Einstein equation near $\mathcal{N}$. Restricted to $r=0$ this implies the following set of geometrical equations on $H$:
\begin{eqnarray}
\label{einsteinrab} \hat{R}_{ab} &=& \tfrac{1}{2} \hat{h}_a \hat{h}_b - \hat{\nabla}_{(a} \hat{h}_{b)} + \Lambda \hat{\gamma}_{ab} + \tr \left(  \hat{E}^2 +\hat{G}^2 \right)\hat{\gamma}_{ab} \\
\label{einsteinf} \hat{F} &=& \tfrac{1}{2} \hat{h}_a \hat{h}^a - \tfrac{1}{2} \hat{\nabla}_a \hat{h}^a + \Lambda - \tr \left( \hat{E}^2 + \hat{G}^2 \right)
\end{eqnarray}
where $\hat{F} \equiv F|_{r=0},\hat{h}_a \equiv h_a|_{r=0}$ are the components of a smooth function and 1-form on $H$, whereas $\hat{\gamma}_{ab} \equiv \gamma_{ab}|_{r=0}$ is Riemannian metric on $H$ with Ricci curvature $\hat{R}_{ab}$ and metric connection $\hat{\nabla}_a$. Observe, that as is typical of degenerate horizons, the equations (\ref{einsteinrab}, \ref{einsteinf}) only contain quantities which are {\it intrinsic} to the horizon.
 
We now consider the Yang-Mills equation (\ref{ymfieldeqn}) near $\mathcal{N}$. We find that restricted to $r=0$ it implies the following equation on $H$:\footnote{This is the $vra$ component of (\ref{ymfieldeqn}) at $r=0$ and degeneracy of $\mathcal{N}$ has been used. The $vab$ component is trivial at $r=0$, whereas the $rab$ component determines the higher order quantity $\hat{\mathcal{W}}_{,rr}$ in terms of $\hat{\mathcal{A}}_{a,r}, \hat{E}$.}
\be
\label{horizonym}
 \hat{\mathcal{D}}_a \hat{G} - h_a \hat{G} - \hat{\epsilon}_{a}^{~b}( \hat{\mathcal{D}}_b \hat{E} - \hat{h}_b \hat{E}) + 2[ \hat{\mathcal{W}}, \hat{\epsilon}_a^{~b} \hat{\mathcal{A}}_{b,r}]=0
\ee
where $\hat{\epsilon}_{ab}$ is the volume form of $\hat{\gamma}_{ab}$ and $\hat{\mathcal{A}}_{a,r} \equiv (\partial_r \mathcal{A}_a)_{r=0}$. 
This horizon Yang-Mills equation reduces to the previously obtained horizon Maxwell equation, see e.g.~\cite{Kunduri:2008tk}.  However,  (\ref{horizonym}) is not just a gauge-covariant version of the Maxwell equation~\cite{Kunduri:2008tk}; it includes a new type of term $[\hat{\mathcal{W}}, \hat{\mathcal{A}}_{,r}  ]$ which encodes information {\it extrinsic} to the horizon. Hence, if $\hat{\mathcal{W}} \neq 0$, a priori it is unclear if the Yang-Mills field on the horizon is constrained as in the abelian case. If fact, for the class of Lie algebras we are considering we may argue this extra term always vanishes.

If $\hat{\mathcal{W}} \neq 0$ at some point on $H$, then (\ref{DW}) implies there is a gauge such that $\partial_a \hat{\mathcal{W}} =0$ and $[\hat{\mathcal{A}}_a, \hat{\mathcal{W}}]=0$. Thus $w \equiv \hat{\mathcal{W}}$ is a fixed element of $\mathfrak{g}$ and by conjugation (i.e constant gauge transformation), we may always assume that $w \in \mathfrak{h}$ where $\mathfrak{h}$ is a Cartan subalgebra. From (\ref{Wcomm}) we deduce that the fields $\hat{\mathcal{A}}_a, \hat{G}, \hat{E}$ are all in the centralizer of $w$ which we denote by $Z_w$. The horizon Yang-Mills equation (\ref{horizonym}) now implies that $[\hat{\mathcal{W}}, \hat{\mathcal{A}}_{a,r} ] \in Z_w$. Semi-simplicity of ad$_w$ then implies that $\hat{\mathcal{A}}_{a,r} \in Z_w$ and hence $[\hat{\mathcal{W}}, \hat{\mathcal{A}}_{a,r} ] =0$ after all.\footnote{If $w$ is a regular element, so $Z_w=\mathfrak{h}$, then the Yang-Mills field is equivalent  to rank($\mathfrak{g}$) Maxwell fields. However, $w$ could also belong to non-abelian centralizers.} Therefore, even if $\hat{\mathcal{W}} \neq 0$, the horizon Yang-Mills equation now simplifies to
\be
\label{nhym}
\hat{\mathcal{D}} \hat{G} - \hat{h} \hat{G} = \hat{\star}_2 ( \hat{\mathcal{D}}\hat{E} - \hat{h} \hat{E} )  \; .
\ee

As in the Einstein-Maxwell case these horizon equations can now be thought of as the full Einstein-Yang-Mills equations for the near-horizon geometry defined above. The limit of the metric is:
\be
g_{NH}= 2\, \td v \left( \td r + r \hat{h}_a(x) \td x^a+ \tfrac{1}{2} r^2 \hat{F}(x)\td v \right) + \hat{\gamma}_{ab}(x) \td x^a \td x^b \; .
\ee
The field strength also always\footnote{Recall we are working in a gauge where $\mathcal{F}$ is independent of $v$. If one does not pick this gauge the near-horizon limit of $\mathcal{F}$ cannot be even defined.}  admits a near-horizon limit due to (\ref{DW}):
\bea
\mathcal{F}_{NH} &=& \hat{E }(x) \td r \wedge \td v - r \hat{\mathcal{D}}_{a} \hat{E} \,\td v \wedge \td x^a + \tfrac{1}{2} \hat{G}(x) \,\hat{\epsilon}_{ab} \td x^a \wedge \td x^b \label{nhF}  \; .
\eea
However, as mentioned above, the gauge field only admits a near-horizon limit if $\hat{\mathcal{W}}\equiv 0$; in this case it given by 
\be
\mathcal{A}_{NH} =\hat{E}(x) r \td v + \hat{\mathcal{A}}_a(x) \td x^a.  \label{ANH}
\ee  
Then, the near-horizon metric and near-horizon gauge field $(g_{NH}, \mathcal{A}_{NH})$ must satisfy the Einstein-Yang-Mills equations (\ref{einsteinequation}), (\ref{ymfieldeqn}).
Indeed one can check directly that the Einstein-Yang-Mills equations for $(g_{NH}, \mathcal{A}_{NH})$ are equivalent to (\ref{einsteinrab}) and (\ref{einsteinf}) and (\ref{nhym}). Furthermore,  since we have argued that $\hat{\mathcal{W}}$ does not actually appear in the horizon equations even when $\hat{\mathcal{W}} \neq 0$, if we simply {\it define} the near-horizon gauge field by (\ref{ANH}) we can still think of the near-horizon geometry as a solution to the Einstein-Yang-Mills equations.

To summarise, the Einstein-Yang-Mills equations for a near-horizon geometry are equivalent to the set of equations (\ref{einsteinrab}), (\ref{einsteinf}) and (\ref{nhym}) for the near-horizon data $(\hat{\gamma}_{ab}, \hat{h}_a, \hat{F}, \hat{E}, \hat{G})$ which are all purely defined on $H$. These equations have inherited the original gauge invariance restricted to $H$; this acts as $(\hat{E}, \hat{G}) \mapsto \hat{U} (\hat{E}, \hat{G}) \hat{U}^{-1}$ and $\hat{\mathcal{A}} \mapsto \hat{U} \hat{\mathcal{A}} \hat{U}^{-1} - \td\hat{U} \hat{U}^{-1}$, where $\hat{U}$ is a group-valued function on $H$.
We will consider the classification of solutions to this system of equations on a compact manifold $H$.  We will focus on $\Lambda \leq 0$, although all of our local results remain valid for $\Lambda>0$.

Before moving on, we note that the contracted Bianchi identity for the horizon metric $\hat{\gamma}_{ab}$ can be written in the useful form
\begin{equation}
\hat{\nabla}_a \hat{F} = \hat{F}\hat{h}_a + 2\hat{h}^b \hat{\nabla}_{[a} \hat{h}_{b]} -\hat{\nabla}^b \hat{\nabla}_{[a} \hat{h}_{b]} - 2 \tr \,  [(\hat{G} \hat{\epsilon}_{ab} + \hat{E} \hat{\gamma}_{ab} ) ( \hat{\mathcal{D}}^b \hat{E} - \hat{h}^b \hat{E})] \label{bianchi} \enskip \; ,
\end{equation}
where we have used (\ref{einsteinrab}), (\ref{einsteinf}) and (\ref{nhym}). 
Henceforth, we will deal with quantities purely defined on $H$ and will drop the ``hats".

\section{Static near-horizon geometries}

Any static black hole must have a static near-horizon geometry, that is $K \wedge \td K=0$ everywhere (not just on $H$). This is equivalent to $\td h=0$ and $\td F=h F$ on $H$. These conditions are solved by $h= \td \lambda$ and $F=F_0 e^\lambda$ for some constant $F_0$ and are sufficient to show that the near-horizon geometry is locally warped product of AdS$_2$ and $H$~\cite{Kunduri:2007vf}.  To solve explicitly for the geometry, we may use the same method as in the Einstein-Maxwell case~\cite{Kunduri:2008tk}. We note that the results of this section generalise those found in~\cite{Hajicek:1982ki, Bicak:1994tq}.

Staticity implies Ricci staticity $K \wedge R(K)=0$, where $R(K)_\mu = R_{\mu\nu} K^\nu$. Using Einstein's equation for a near-horizon geometry Ricci staticity implies that ${\mathcal{D}_a}E = h_a E$ on $H$. Hence ${\mathcal{D}} (e^{-\lambda} E) = 0$ and thus $q^2 \equiv e^{-2\lambda}  \tr \, E^2$ is a constant on $H$. The horizon Yang-Mills equation (\ref{nhym}) reduces to ${\mathcal{D}}_a G = h_a G$ and hence ${\mathcal{D}} (e^{-\lambda} G) = 0$, so we learn that $p^2 \equiv e^{-2\lambda} \tr \, G^2$ is also a constant on $H$. 

So far the analysis has been local; it applies to any coordinate patch $U_i$ such that $h=\td\lambda_i $. Since $\tr \, E^2$ and $\tr \, G^2$ are invariants of the solution from the above we deduce that on any overlap $U_i  \cap U_j$ we must have $q_i^2 e^{2\lambda_i} = q_j^2 e^{2\lambda_j}$ and $p_i^2 e^{2\lambda_i} = p_j^2 e^{2\lambda_j}$; for a non-trivial Yang-Mills field we see that either all the $q_i$ are non-zero or all the $p_i$ are non-zero. Since the $\lambda_i$ in each $U_i$ are defined only up to an additive constant we may always arrange $q_i=q_j$ or $p_i=p_j$ and hence $\lambda_i=\lambda_j$. Therefore there exists a globally defined function $\lambda$ such that $h = \td\lambda$ irrespective of the topology of $H$ (of course if $H=S^2$ this is automatic).

Observe that the source terms  in the horizon Einstein equations $\tr ( E^2 +G^2 ) = (q^2+p^2)e^{2\lambda} $ are of the same form as in the Einstein-Maxwell case.  If $\lambda$ non-constant, one can use the same method as in~\cite{Kunduri:2008tk} to explictly solve for the horizon metric $\gamma_{ab}$ and show that it  can not be extended smoothly onto a compact $H$ (at least for $\Lambda\leq 0$).
Hence compactness requires $h=\td\lambda \equiv 0$, which implies $E$ and $G$ are covariantly constant: thus $\tr \, E^2, \tr \, G^2$ must be constants. The horizon equations now reduce to 
\be
{R}_{ab} = (\Lambda + \tr \,( E^2 +G^2)) \gamma_{ab} \; , \qquad \qquad F= \Lambda - \tr \,( E^2 +G^2) \; ,
\ee
so that $H$ is a constant curvature space and $F=F_0$ is a constant. The near-horizon geometry is simply the direct product of a 2d Lorenzian maximally symmetric space and a 2d constant curvature space. For $F<0$, as must be the case if $\Lambda \leq 0$, it is AdS$_2 \times H$ with $H=S^2, T^2, \Sigma_g$ depending on the sign of curvature, where $\Sigma_g$ is a Riemann surface of genus $g$ (only $S^2$ is allowed for $\Lambda =0$). 

If $E \equiv 0$ the problem reduces to solving $\mathcal{D}_aG=0$ on $H$, which is equivalent classifying Yang-Mills connections on $S^2$ and more generally on a Riemann surface of higher genus, a problem which has been solved~\cite{Atiyah:1982fa}. In particular, for $S^2$  the moduli space Yang-Mills connections is in one-to-one correspondence with conjugacy classes of closed geodesics on the gauge group~\cite{Atiyah:1982fa, Friedrich:1985vr}. We deduce that for $SU(2)$ gauge group all solutions on $S^2$ must be abelian. For more general gauge group it may be interesting to construct explicit non-abelian solutions, although we will not pursue this here.

Finally consider the case where $E \neq 0$ at least at a point.  We have that $0={\mathcal{D}}^2 E=[{\mathcal{F}}, E]$ and hence $[E, G]=0$. Furthermore, since $E$ is covariantly constant we may choose a gauge such that it is a constant on $H$; then $[\mathcal{A}_a, E]=0$. It follows that $\mathcal{A}_a, G \in Z_E$. By a constant conjugation we may assume $E \in \mathfrak{h}$ is in a Cartan subalgebra. If $E$ is a regular element of $\mathfrak{g}$ then all fields are in $\mathfrak{h}$ and hence the system is equivalent to rank($\mathfrak{g}$) Maxwell fields. For $SU(2)$ gauge group this is the only possibility so in this case there are no non-abelian solutions.  For more general gauge group, if $E$ is a singular element then the centralizer $Z_E$ is non-abelian and the problem reduces again to the 2d Yang-Mills equations on a Riemann surface with a gauge group broken to the centralizer of $E$. 

\section{Axisymmetric near-horizon geometries}

Motivated by the rigidity theorem for rotating black holes, we will now assume the spacetime and extreme horizon are axisymmetric. That is, we assume  there exists a $U(1)$ isometry which commutes with the $\mathbb{R}$ isometry generated by $K$ (and hence leaves the horizon $\mathcal{N}$ invariant). We denote the corresponding Killing field by $m$ and we assume the spacetime gauge field is also invariant up to gauge transformations, so $\mathcal{L}_m {\mathcal{A}}_\mu =  {\mathcal{D}}_\mu \mathcal{V}_m$ for a group-valued function $\mathcal{V}_m$. 

The vector field $m$ must be tangent to $H$ and hence generates a $U(1)$ action on $H$; it  restricts to a Killing field of the metric ${\gamma}_{ab}$ on $H$ which also leaves the rest of the near-horizon data ${F}, {h}_a$ invariant. The near-horizon gauge field data inherits the following invariance properties $\mathcal{L}_m {\mathcal{A}_a} = {\mathcal{D}_a} {\mathcal{V}}_m$, $\mathcal{L}_m {G} = [{G},  {\mathcal{V}}_m]$, $\mathcal{L}_m {E} = [ E, {\mathcal{V}}_m ]$, where $\mathcal{V}_m$ is now a function on $H$.

The existence of a $U(1)$-action on $H$ constraints its topology: if the action is free it must be $T^2$, otherwise it must be $S^2$ in which case there are exactly two fixed points (the poles). We will first consider the $S^2$ case. Now consider the closed 1-form on $H$ defined by $i_m {\epsilon}$. It follows there exists a function $x$ such that $\td x =i_m {\epsilon}$. Compactness implies that there exists a global maximum and minimum for $x$, so $x_1 \leq x \leq x_2$. Since $(\td x)^2 = | m|^2$ we see that $x$ can be used as a coordinate at any point where $m \neq 0$. We deduce that the fixed points of $m$ correspond to the endpoints $x=x_1,x_2$. 
Therefore we can introduce coordinates $(x, \phi)$ for $x_1<x<x_2$ such that $m= \partial / \partial \phi$, in which the near-horizon metric can be parameterised as
\be
\gamma_{ab} \td x^a \td x^b = \frac{\td x^2}{B(x)} + B(x) \td\phi^2, \qquad \qquad h_a \td x^a = \Gamma(x)^{-1}(B k(x) \td\phi -\Gamma'(x) \td x) \; ,
\ee
where $B(x)>0$ and $B(x_1)=B(x_2)=0$ and $\Gamma(x)>0$ everywhere. Smoothness requires the absence of conical singularities at the end points $x=x_1, x_2$: this is equivalent to $B'(x_1)=-B'(x_2)=2$ and $\phi \sim \phi+2\pi$.

Now consider the gauge field. We may choose a gauge such that $\mathcal{V}_m=0$ and hence $\partial_\phi {\mathcal{A}}_a = \partial_\phi E= \partial_\phi G= 0$, i.e. $\mathcal{A}_a, E, G$ are only functions of $x$.  Furthermore by a residual axisymmetric gauge transformation we can also set ${\mathcal{A}}_x=0$. 
In this gauge the horizon gauge field is simply
\be
{\mathcal{A}}_a \td x^a = a(x) \td\phi 
\ee
where $a \equiv i_m \mathcal{A}$ is a $\mathfrak{g}$-valued function on $H$. It follows that
\be
 G(x) = a'(x) \; .
 \ee
The horizon Yang-Mills equations (\ref{nhym}) now reduce to the coupled ODE system
\bea
B (\Gamma G)'+ B k E &=& \Gamma [ a, E]
\label{YM1}\\
B(\Gamma E)' - Bk G &=& -\Gamma [a,G] 
 \label{YM2}  \; .
\eea

Now consider the $x \phi$ component of (\ref{einsteinrab}). The terms from the Yang-Mills fields do not contribute and one finds as in the vacuum case $k'=0$, so $k$ must be a constant. If $k=0$ the near-horizon geometry is in fact static;  as shown in the previous section static near-horizon geometries can be treated more generally without the assumption of axisymmetry.

The $x$ component of (\ref{bianchi}) can be simplified using $k'=0$ and (\ref{YM2}), resulting in the expression
\be
B A'= 4\Gamma \, \tr( E [ a, G] )   \; ,   \label{xbianchi}
\ee
where we have defined the function
\be
A \equiv \Gamma F- k^2 \Gamma^{-1}B  \; .
\ee 
The significance of this quantity is revealed by changing $r \to \Gamma(x) r$ in the full near horizon geometry, which results in
\be
g_{NH} = \Gamma(x) [ A r^2 \td v^2 +2\td v \td r ] + \frac{\td x^2}{B(x)} + B(x) (\td\phi+ k r\td v)^2  \; .
\ee
In an abelian theory, such as Einstein-Maxwell theory, the righthand side of (\ref{xbianchi}) must vanish. In that case $A$ is a constant which can be shown to be negative for $\Lambda \leq 0$; then the metric in the square brackets is AdS$_2$ and the near-horizon geometry inherits all its isometries (since $k$ is constant).  The non-abelian structure of Einstein-Yang-Mills theory thus appears to obstruct this symmetry enhancement phenomena.

Let us now study the obstruction term
\be
T \equiv \tr( \Gamma E [ a, \Gamma G] )
\ee
where the extra factors of $\Gamma$ appear for convenience. Let us also define
\be
S \equiv \Gamma^2 \tr ( E^2 + G^2)  \; .
\ee  
These quantities can be constrained using the Yang-Mills equations. Indeed equations (\ref{YM1}), (\ref{YM2}) allow one to establish the crucial identities
\bea
B S' &=& -4 T  \label{id1}  \\
B T' &=& - \Gamma^2 \tr( [a, G]^2+ [a, E ]^2 ) \; .  \label{id2}
\eea
We may use these identities together with a global argument on $H$ as follows.

First note that the vector field $X \equiv B \partial / \partial x$ is globally defined on $S^2$ and vanishes at $x=x_1,x_2$. Hence given any smooth function $f$ on $H$, the function $X(f)$ must also be smooth everywhere on $H$ and also vanishes at $x=x_1,x_2$.  It is clear that $S$ is invariantly defined on $H$; hence (\ref{id1}) implies $T$ is smooth on $H$ and vanishes at the endpoints
\be
T(x_1)=T(x_2)=0.  \label{bcT}
\ee
It then follows from (\ref{id2}) that $X(T) \leq 0$ and $X(T)|_{x=x_1,x_2}=0$. Assume there is a single point in the open interval $x_1<x<x_2$ such that $X(T)<0$. At this point $T'<0$ and therefore
\be
T(x_2)-T(x_1) =\int_{x_1}^{x_2} \td x \;T' <0  \; .
\ee
This clearly contradicts (\ref{bcT}) and therefore we deduce that $T \equiv 0$ for all $x_1<x<x_2$.  Hence we have shown that the obstruction term in (\ref{xbianchi}) vanishes and deduce that
\be
A(x)= A_0
\ee
where $A_0$ is a constant. The sign of $A_0$ can be determined using (\ref{einsteinf}) which gives
\be
A_0 = \frac{1}{2} {\nabla}^2 \Gamma - \frac{k^2 B}{2\Gamma} + \Gamma \Lambda - \frac{S}{\Gamma} \; . \label{A0} 
\ee
By integrating this equation over $H$ we deduce that for $\Lambda \leq 0$ a non-trivial solution (i.e. either $k\neq 0$ or $S \neq 0$) must have $A_0 <0$.  By the above remarks this shows the near-horizon geometry possesses the AdS$_2$ symmetry enhancement as in the abelian theory. 

Observe that from (\ref{id1}) and (\ref{id2}) the condition $T=0$ allows us to deduce that $S=S_0$ is a constant and $[a, G]=[a, E] \equiv 0$.  Commuting (\ref{YM2}) with $a$ then shows that $[a, E']=0$, which can be used to deduce $[G, E]=0$. Then commuting (\ref{YM2}) with $E$ and $G$, shows $[E, E']=0$ and $[G,E']=0$ respectively.  This shows that all the components of near-horizon gauge field and field strength (\ref{nhF}) commute.

The classification problem now essentially reduces to that in Einstein-Maxwell theory  
which has been previously solved, so we will be brief (although the present argument is more efficient than in~\cite{Kunduri:2008tk}). Take the $xx$ component of (\ref{einsteinrab}) and subtract $B^{-2}$ times the $\phi\phi$ component of (\ref{einsteinrab}) to get
\be
\Gamma''- \frac{\Gamma'^2}{2\Gamma} - \frac{k^2}{2\Gamma} = 0  \; .  \label{Gammaeq}
\ee
If $\Gamma$ is a constant then $k=0$ and hence we recover the static case $h_a \equiv 0$.  If $\Gamma$ is non-constant  then this equation for $\Gamma$ can be used to rewrite (\ref{A0}) as
\be
\left( \frac{B\Gamma}{\Gamma'} \right)'= \frac{2( A_0\Gamma -\Lambda \Gamma^2+S_0)}{\Gamma'^2} \; .  \label{Beq}
\ee
The general solution to (\ref{Gammaeq}) is
\be
\label{Gam}
\Gamma=\frac{k^2}{\beta}+ \frac{\beta x^2}{4}
\ee
where $\beta>0$ is an integration constant and we have used the shift freedom in the definition of $x$ to fix the other constant. This can then used to integrate for $B$ using (\ref{Beq})
\be
B(x)= \frac{P(x)}{\Gamma}
\ee
where $P$ is a polynomial given by
\be
P(x) =-\frac{\beta \Lambda x^4}{12}+(A_0-2\Lambda k^2 \beta^{-1})x^2+c_1 x -\frac{4k^2}{\beta^2}(A_0- \Lambda k^2\beta^{-1}) - \frac{4S_0}{\beta}
\ee
and $c_1$ is an integration constant. We have thus completely solved for the metric and a global analysis reveals that the horizon metric extends to smoothly onto $S^2$ if and only if $c_1=0$ (at least for $\Lambda \leq 0$). 

We now turn to the Yang-Mills equations (\ref{YM1}) and (\ref{YM2}) which reduce to:
\bea
(\Gamma G)'+ k E &=& 0\\
(\Gamma E)' - k G &=& 0   \; .
\eea
By expanding in any Lie algebra basis the components of $(E,G)$ each satisfy the same equations as in the Maxwell case. Assuming $k \neq 0$ one finds
\be
E = \frac{x q - \left( \frac{k^2}{\beta} - \frac{\beta x^2}{4} \right) p }{\Gamma^2}, \qquad  a = a_0+ \frac{x q - \left( \frac{k^2}{\beta} - \frac{\beta x^2}{4} \right) p }{k\Gamma}
\ee
where $a_0, q, p$ are fixed elements in $\mathfrak{g}$ and recall $G=a'$. Since these fields and their first derivatives must commute for all $x$, we deduce that $a_0, q, p$ all commute.

The above shows that the most general axisymmetric near-horizon geometry and gauge field with $H=S^2$ is isometric to that of the abelian embedded extreme Kerr-Newman black with a cosmological constant (see~\cite{Kunduri:2008tk} to deduce the explicit coordinate and parameter change). 

We close by considering toroidal horizon topology $H=T^2$. In this case one can again introduce coordinates $(x, \phi)$, this time both periodic, such that the horizon metric take the same form with $B(x)>0$ everywhere, but instead $h$ may have an extra term of the form $c B(x)^{-1} \td x$ for some constant $c$, see e.g.~\cite{Holland:2010bd}. The $x\phi$ component of (\ref{einsteinrab}) now implies $k' =c\, k/B$ which integrates to $k(x) = k_0 \exp(c \int^x_{x_0} B(x)^{-1} \td x)$. Since the integrand in the exponent is positive this means that $k(x)$ is a monotonic function which is in contradiction to the fact that $k$ must be a periodic function of $x$. Hence for $k\neq 0$ we must have $c=0$ after all. Therefore the horizon equations in the toroidal case are identical to the $S^2$ case. From above we saw that if $k \neq 0$ then $\Gamma(x)$ is given by  (\ref{Gam});  but since $\Gamma$ is a globally defined function on $H$ it must be periodic in $x$ and hence we have a contradiction. This shows there are no axisymmetric near-horizon geometries  which are non-static (i.e. $k \neq 0$) and have $H=T^2$. Note that this proof is equally valid in pure Einstein-Maxwell theory or even pure gravity; for $\Lambda<0$ this fact does not seem to have been shown before (for $\Lambda \geq 0$ it is easily follows by integrating the trace of the general horizon equation (\ref{einsteinrab})).

The above results thus completely classify all axisymmetric near-horizon geometries with a compact horizon section in Einstein-Yang-Mills theory with a cosmological constant. We conclude that the near-horizon uniqueness present in Einstein-Maxwell theory persists in the non-abelian Einstein-Yang-Mills theory.
\\

\noindent {\bf Acknowledgements:} CL is supported by a Principal Career Development Scholarship at the University of Edinburgh. JL is supported by an EPSRC Career Acceleration Fellowship. We would like to thank Jos\'e Figueroa-O'Farrill for useful discussions.

\end{document}